\documentclass[12pt,a4paper]{article}
\pdfoutput=1

\usepackage{bbding}
\usepackage{pifont}
\usepackage{wrapfig}
\usepackage{amsmath}
\usepackage{verbatim}
\usepackage{amssymb}
\usepackage{inputenc}
\usepackage{textcomp}
\usepackage{appendix}
\usepackage{mathrsfs}
\usepackage{float}
\usepackage{xfrac}
\restylefloat{table}
\usepackage{subfig}
\usepackage{mathtools}
\usepackage{tabularx}
\usepackage{bigints}
\usepackage{wasysym}
\usepackage[normalem]{ulem}
\usepackage{graphicx}
\usepackage{appendix}
\usepackage{amsfonts,setspace}
\usepackage{xcolor}
\usepackage{comment}
\usepackage{caption}
\usepackage{hyperref}
\usepackage{cite}

\newcommand{\e}{\epsilon}

\newcommand{\be}[1]{\begin{equation}\label{#1} }
\newcommand{\ee}{\end{equation}}
\newcommand{\bea}[1]{\begin{eqnarray}\label{#1} }
\newcommand{\eea}{\end{eqnarray}}
\newcommand{\p}{\partial}

\renewcommand{\L}{\mathcal{L}}

\renewcommand{\a}{\alpha}

\renewcommand{\b}{\beta}
\renewcommand{\t}{\tau}

\newcommand{\bes}{\begin{subequations}}
\newcommand{\ees}{\end{subequations}}

\newcommand{\lb}{\left[}
\newcommand{\rb}{\right]}
\newcommand{\fE}{f^{\mu\nu}_{(E)}}
\newcommand{\fEE}{f_{\mu\nu}^{(E)}}
\newcommand{\fM}{f_{\mu\nu}^{(M)}}
\newcommand{\fMM}{f^{\mu\nu}_{(M)}}
\usepackage[margin=0.75in]{geometry}
\hypersetup{colorlinks=false, linkcolor=blue, citecolor=red}
\newcommand{\vs}[1]{\vspace{#1 mm}}
\begin{document}

	\begin{flushright}
	\end{flushright}
	
\begin{center}
		{\Huge{\bf ModMax meets GCA}}\\

\vs{10}

{\large
Aritra Banerjee${}^{a,\,}$\footnote{\url{aritra.banerjee@oist.jp}}, Aditya Mehra${}^{b,c\,}$\footnote{\url{aditya.mehra@ed.ac.uk}}, \\ 

\vskip 0.3in

{\it ${}^{a}$ Okinawa Institute of Science and Technology, \\1919-1 Tancha, Onna-son, Okinawa 904-0495, JAPAN }\vskip .5mm

{\it ${}^{b}$Department of Physics, BITS-Pilani, K K Birla Goa Campus,\\ Zuarinagar, Goa-403726, INDIA}\vskip .5mm

{\it ${}^{c}$School of Mathematics and Maxwell Institute for Mathematical Sciences,\\ University of Edinburgh, Peter Guthrie Tait Road, \\Edinburgh EH9 3FD, UK}\vskip.5mm}

\end{center}

\vskip 0.35in

\begin{abstract}
A maximally symmetric non-linear extension of Maxwell’s theory in four dimensions called ModMax has been recently introduced in the literature. This theory preserves both electromagnetic duality and conformal invariance of the linear theory. In this short paper, we introduce a Galilean cousin of the ModMax theory, written in a covariant formalism, that is explicitly shown to be invariant under Galilean Conformal Symmetries. We discuss the construction of such a theory involving Galilean electromagnetic invariants, and show how the classical structure of the theory is invariant under the action of Galilean Conformal Algebra (GCA).
	\end{abstract}

	\newpage

\tableofcontents

\section{Introduction}

Maxwell's electrodynamics in $4d$ is special, in the sense it has two very important symmetries, namely the four-dimensional conformal symmetry and electric-magnetic duality. Maxwell theory  is certainly the most successful and well known Gauge Theory of $U(1)$ fields since its introduction one and half century ago. Most notably Maxwell's equations derived from this theory are linear in the field strength $F_{\mu\nu}$. It has been widely accepted that a generic quantum theory of Electrodynamics should have higher order corrections to the linear terms in $F_{\mu\nu}$ in the Lagrangian, arising from loop contributions \cite{Heisenberg:1936nmg}, which reduces to the pure Maxwell term in the low energy limit. This gave rise to the question of whether there exists other classical Lagrangians for $U(1)$ gauge fields which are already non-linear in  the field strength and gives rise to Maxwell theory in an effective description. These theories may capture new physics at different energy scales where the full non-linear theory has to be taken care of, leading to corrections to known results. Thus materialised the studies of Non-Linear Electrodynamics (NLED), which has been going strong for more than a century already.

\medskip

Notable examples of NLEDs include the famous Born-Infeld theory \cite{Born:1934gh}, a crucial component of String Theory in the study of D-branes \cite{Polchinski:1995mt}, that makes sure to keep self energy of point particles finite. Other well studied examples include actions involving various functionals of the field tensor and a nice review for these constructions can be found in \cite{Sorokin:2021tge} and references therein. But the main caveat lies in the problem that a generic NLED in $4d$ is not conformally invariant, neither is it invariant under Hodge duality rotations. So the question people have been asking for decades, reads, \textit{is it possible to write down a Lorentz invariant non-linear theory of Electrodynamics that preserves the symmetries of Maxwell Lagrangian?} As far as conformal invariance is concerned, it has been shown that as long as the lagrangian is a homogeneous function of degree one of Maxwell Lorentz invariants, the theory remains invariant. However, the requirement of electromagnetic duality invariance takes a more involved form, as shown first by Bialynicki-Birula \cite{Bialynicki-Birula:1984daz}. Reconciling the two conditions seemed to be an involved problem for decades, for example, Born-Infeld theory gives rise to manifestly duality invariant equations, but it is not conformal invariant. 

\medskip

Only recently, the question we posed have been completely answered by Bandos, Lechner, Sorokin, and Townsend  \cite{Bandos:2020jsw}  (see also \cite{Kosyakov:2020wxv}) who proposed a simultaneously duality-invariant and conformal theory of $U(1)$ fields, that reduces to Maxwell theory in a zero coupling limit. This theory has been generically called Modified Maxwell theory or by the nickname ``ModMax''. This has generated considerable interest in last couple years and have been shown to have many interesting properties as seen in the classical solutions \cite{Dassy:2021ulu}, hamiltonian formulation \cite{Escobar:2021mpx,Bandos:2020hgy}, coupling to charged particles \cite{Lechner:2022qhb}, supersymmetric and other generalizations \cite{Kuzenko:2021xxx, Bandos:2021rqy,Kuzenko:2021qcx, Avetisyan:2021heg} and connection to black hole solutions \cite{Flores-Alfonso:2020euz,BallonBordo:2020jtw,Bokulic:2021dtz,Kruglov:2022qag,Barrientos:2022bzm}. It has also been shown that ModMax theories can be generated by a $T\overline{T}$-like (or $\sqrt{T\overline{T}}$ like) deformation of Maxwell theories \cite{Babaei-Aghbolagh:2022uij,Ferko:2022iru,Conti:2022egv, Ferko:2022lol}, and a string theory context for ModMax has been introduced in \cite{Nastase:2021uvc}. This list of course does not do justice to the literature, and readers are directed to references and citations of these papers. Certainly, the general symmetry structure of ModMax is intriguing on its own and will be part of various studies in the near future.  

\medskip

Our goal in this note, however, is to meander from the well traversed paths and try something very new. We would like to focus on the conformal nature of the ModMax theory, and would like to see what happens when one looks at these symmetries going away from the relativistic situation. Specifically, we set out trying to write an analogue \textit{Galilean Covariant} nonlinear Lagrangian which is invariant under the $4d$ Galilean Conformal Algebra (GCA).  These symmetries arise when the we take a non-relativistic (speed of light going to infinity) limit on the $d$ dimensional conformal algebra. At each and every dimension, this limit results in an infinite-dimensional Galilean conformal algebra \cite{Bagchi:2009my}, in contrast to their relativistic cousins which are only infinite dimensional in $2d$ \cite{Belavin:1984vu}. Galilean electrodynamics has been studied for a long time, starting as early as with Le Ballac and Levy-Leblond \cite{LBLL}. In recent years theories of Galilean Electrodynamics have generated newfound interest due to the larger and rich symmetry structures associated to it, and in \cite{Bagchi:2014ysa} a reformulation of Galilean Conformal Electrodynamics in various dimensions was introduced via taking non-relativistic (NR) limits on the equations of motion of the Maxwell theory, and has subsequently been developed in a bunch of works \cite{Duval:2014uoa, Festuccia:2016caf, Bergshoeff:2015sic, Bleeken:2015ykr, Banerjee:2019axy, Hansen:2020pqs, Chapman:2020vtn,Banerjee:2022uqj}. A caveat for taking such a limit is loss of manifest electric-magnetic duality, as the physics splits in two sub-sectors, where either the electric or the magnetic components of the gauge field $A_\mu$ dominates. The other problem is, the procedure of taking NR limits does not work accurately on the action formalism and hence a proper covariant Galilean electrodynamic action is hard to write down. 

\medskip 

The search for a covariant Galilean Electrodynamics theory needs to be addressed by putting the gauge fields explicitly on a Non-Lorentzian manifold, in this case a Newton-Cartan manifold. Newton-Cartan structures arise when we take the speed of light to infinity and the usual Riemannian notion of a manifold degenerates, paving the way for Galilean Relativity \cite{Duval:1993pe, Duval:2009vt,Duval:2014uoa}. These manifolds in general have a fibre bundle structure that keeps temporal and spatial diffeomorphisms separate from each other. An attempt to write down a Galilean Covariant electrodynamics Lagrangian was recently made in \cite{Mehra:2021sfx}, one which simultaneously describes electric and magnetic dominated realms of the theory. This Galilean Maxwell Lagrangian will be the building block of our current work, and since that Lagrangian is manifestly invariant under the $4d$ GCA, we will try to introduce an explicit non-linear Covariant Lagrangian with a ModMax-like form, consequently showing the GCA invariance for the same as well. This will be the first instance of a nonlinear covariant Galilean Conformal Electrodynamics (NLGCED) Lagrangian in the literature, to the best of our knowledge. 

\medskip
 
 The rest of this short paper is organised in the following way:  in section \ref{sec2} we will briefly review the structure of ModMax electrodynamics. In section \ref{sec3} we will revisit Newton-Cartan structures and construction of a gauge field Lagrangian on such a structure. Here we will slightly differ from the approach of \cite{Mehra:2021sfx}, and instead focus on the transformation of components of gauge fields under Galilean Conformal symmetries. We will also discuss in detail the structure of Galilean invariant field bilinear that will be important to our construction. Then in section \ref{sec4} we will go ahead and present our Galilean ModMax-like Lagrangian and show its explicit invariance under GCA symmetries, giving the generic form of ModMax Lagrangians (with square roots) an aura of universality when it comes to conformal invariance both in and beyond Lorentzian cases. In section \ref{sec5} we will have further discussions and talk about probable future extensions. 
 \newpage
\section{ModMax as Conformal non-linear Electrodynamics}\label{sec2}
 As mentioned earlier, source-free ModMax theory in $4d$ Minkowski spacetime is both conformally invariant and E-M duality invariant. In this section we will briefly revisit some aspects of the ModMax Lagrangian, which will be later crucial for the discussion of the analogous Galilean theory.

\subsection*{Lagrangian and symmetries of the Relativistic ModMax}
In $4d$, generic electrodynamics Lagrangians can only be functions of the field strength tensor, and does not include the derivatives of those. For $4d$ Maxwell electrodynamics, there are just two Lorentz invariant quantities, i.e.
\be{}
S = -\frac{1}{4}F^{\mu\nu}F_{\mu\nu}= \frac{1}{2}\left(\bf{E}^2 -\bf{B}^2  \right), ~~P = -\frac{1}{4}F^{\mu\nu}\tilde{F}_{\mu\nu} = \bf{E}\cdot\bf{B}.
\ee
Here the first one is a Lorentz scalar and the second one is pseudo-scalar, where the field strength and the Hodge dual field strength is defined as 
\be{}
F_{\mu\nu} = \p_\mu A_\nu-\p_\nu A_\mu,~~~\tilde{F}_{\mu\nu} =\frac{1}{2}\e_{\mu\nu\rho\sigma}F^{\rho\sigma},
\ee 
where $A^\mu$ is an $U(1)$ gauge field. A generic theory of Electrodynamics in this dimension will be an analytic function of these Lorentz invariants and will reduce to the pure Maxwell term in some (often weak field) limit. 

As shown in \cite{Bandos:2020jsw}, the unique one-parameter deformed family of Lorentz invariant modifications of the Maxwell Lagrangian which stays conformal invariant and produces E-M duality invariant equations of motion is the ModMax theory. 
The Lagrangian density for this theory can be written down as
\bea{}\label{MM}
\mathcal{L}_{MM}&=& \cosh\gamma~S + \sinh\gamma~ \sqrt{S^2+P^2}\\ \nonumber
&=&-\frac{\text{cosh}\gamma}{4}\big[F^{\mu\nu}F_{\mu\nu}\big]+\frac{\text{sinh}\gamma}{4}\sqrt{(F^{\mu\nu}F_{\mu\nu})^2+(\tilde{F}^{\mu\nu}F_{\mu\nu})^2}.
\eea
Here $\gamma$ is the dimensionless parameter that controls the deformation, and at $\gamma=0$ the theory reduces down to pure Maxwell, with $\gamma>0$ having a well defined solution space. The conditions posed by causality and unitarity demand the coupling constant to be non-negative $\gamma \geq 0$ \cite{Bandos:2020jsw}. The square root term can be thought of as a deformation to Maxwell theory, and in general is proportional to $\left(\bf{E}^2 +\bf{B}^2  \right)$, i.e. the Hamiltonian of the pure Maxwell theory. The term under square root can also be rewritten as $\mathcal{F}^{\mu\nu}\mathcal{\bar{F}}_{\mu\nu}$, where $\mathcal{F}^{\mu\nu}= F^{\mu\nu}+i\tilde{F}^{\mu\nu}$, and the bar denotes complex conjugation. This structure of the deformation turns out to be crucial to show invariance of the equations of motion under conventional  Electromagnetic duality as well.

The unique nature of the ModMax Lagrangian makes sure of conformal invariance as it is a homogeneous function of degree one under scale transformations \cite{Bandos:2020hgy}. One can see this symmetry via explicitly computing the stress-energy tensors of this theory, which turn out to be traceless and explicitly proportional to their Maxwell cousins.

\subsection*{Equations of motion}
The equations of motion of the Lagrangian \eqref{MM} is given by
\bea{eomm}\p_{\mu}\Bigg[(\text{cosh}\gamma)F^{\mu\nu}-\text{sinh}\gamma\Bigg(\frac{(F^{\a\b}F_{\a\b})F^{\mu\nu}+(\tilde{F}^{\a\b}F_{\a\b})\tilde{F}^{\mu\nu}}{\sqrt{(F^{\a\b}F_{\a\b})^2+(\tilde{F}^{\a\b}F_{\a\b})^2}}\Bigg)\Bigg]=0,\eea
which again boils down to the standard equation $dF = 0$ when we put $\gamma=0$. One can note that these equations of motion for generic $\gamma$ are ill-defined for solutions having null electromagnetic fields, i.e. $FF = F\tilde{F} = 0$, like in the case of Electromagnetic waves. It has been shown that these pathologies can be cured if one instead works in the Hamiltonian formalism \cite{Bandos:2020jsw}.

The duality invariance associated to these equations, although not so important for our discussion here, is given by the so called Galliard-Zumino duality conditions \cite{Gaillard:1981rj} for non-linear Electrodynamics, 
\be{tensE}
\tilde{F}^{\mu\nu}F_{\mu\nu}=\tilde{E}^{\mu\nu}E_{\mu\nu},
\ee
which replaces the usual rotations of field strengths well-known for the source free Maxwell equations. Here the excitation field strength is given by $E_{\mu\nu} = \frac{\p\mathcal{L}_{MM}}{\p F^{\mu\nu}} $ and the hodge dual of that is defined in the usual way. Notice that $E^{\mu\nu}$ turns out to be just $F^{\mu\nu}$ for a pure Maxwell theory. So it is clear that equations of motion deduced from general non-linear theories would demand a symmetry under $U(1)$ rotations of $E^{\mu\nu}$ and $\tilde{F}^{\mu\nu}$. \footnote{One can however write down an action principle for generic non-linear electrodynamics that also manifestly shows electromagnetic duality symmetry beyond the equations of motion. See \cite{Avetisyan:2021heg} for such a democratic formulation. } One can notice that our equation of motion \eqref{eomm} can be rewritten using the tensor $E^{\mu\nu}$ in \eqref{tensE} as 
\be{}
\p_\mu E^{\mu\nu} = 0,
\ee
which makes sure this and the usual Bianchi identity $d\tilde{F} = 0$ are a set of duality invariant non-linear equations of motion for the ModMax theory. 


\section{Covariant Galilean Conformal Electrodynamics}\label{sec3}
In this section, we will discuss the formalism associated to a Covariant formulation of Galilean Electrodynamics. As we have introduced before, it is essential to start with a geometric formulation of tensors on a Newton-Cartan manifold and put gauge fields explicitly on it to understand the true nature of Galilean conformal symmetries.

\subsection{Galilean geometry}
In Galilean sense, usual Riemannian metrics are of no use since they are degenerate and can't be used to raise or lower indices on objects. At the limit of $c\to \infty$ the Poincar\'{e} group is replaced by the Galilei group, and the kinematical structure of the group allows one to define a manifold, called the Newton-Cartan manifold. 
The main ingredients of an intrinsically Galilean (or Newton-Cartan) manifold is the degenerate spatial metric $h^{\mu\nu}$ and a choice of null time direction $\t_\mu$ that gives rise to another two-index object $\t_{\mu\nu}=-\t_\mu\t_\nu$  \cite{Kuenzle:1976sdk, Dautcourt:1990sds, Andringa:2010it, Bleeken:2015ykr}. In $4d$, the simplest choice to represent these tensors are:
\bea{ht1}
h^{\mu\nu} = \begin{bmatrix*}[r]
0 & \phantom{-}0 \\
0 & \phantom{-}1_{3\times 3}\\
\end{bmatrix*},
\quad
\tau_{\mu\nu} = \begin{bmatrix*}[r]
-1 & \phantom{-}0\\
0 & \phantom{-}0_{3\times 3}\\
\end{bmatrix*}.
\eea
These two non-invertible Galilean tensors are used to define contravariant and covariant Galilean vectors and the nowhere vanishing time-like vector field $\t_\mu$ is given by,
 \bea{tau}
\tau_\mu =\begin{bmatrix*}[r]
1 & 0 & 0 & 0 
\end{bmatrix*} ~~~~\tau_\mu \tau^\mu =1.
\eea
These two geometric ingredients are orthogonal in the sense 
$h^{\mu\nu}\tau_{\nu}=0$.
Both these objects remain invariant under Galilean boosts and rotation. We can remind the reader that time is absolute in Galilean relativity, which is inherent in these invariant structures. 
For the $(h,\t)$ duo, which defines a gauge choice (i.e. the form of the tensors in \eqref{ht1}) for a Newton-Cartan spacetime, the respective covariant and contravariant objects are given by projective inverses of $(h^{\mu\nu}, \t_\nu)$, and are given by,
\bea{ht}
h_{\mu\nu} = \begin{bmatrix*}[r]
a & \phantom{-}b_i \\
b_i & 1_{3\times 3}
\end{bmatrix*},
\quad
\t^{\mu\nu} = \t^{\mu} \t^{\nu} \;\; \text{where} \; \t^\mu = (1, c_1, c_2, c_3).
\eea
These expressions follow from the definition of the Galilean tensors using projective inverse definitions given by $\t^\mu \t_\mu = 1$ and by $h^{\mu\a}h_{\a\b} h^{\b\nu} = h^{\mu\nu}$. One should note that these projective inverses are not generally Galilean invariants for all choices of the constants $(a,b_i,c_i)$. 

These tensors are crucial in defining Galilean objects in the theory, i.e. a covariant vector $\tilde{K}_\mu$ will be defined from the knowledge of a contravariant vector $K^\mu$ via the operation $\tilde{K}_\mu = \t_{\mu\nu}K^\nu$, and an opposite operation $(K_\mu \to \tilde{K}^\mu)$ will be done via $h^{\mu\nu}$. Contrary to relativistic tensors, these operations are not invertible since temporal and spatial components are split from each other due to the structure of $(h^{\mu\nu}, \t_\nu)$. Similarly, covariant derivatives are defined as $\p_\mu = (\p_t,\p_x,\p_y,\p_z)$, while the associated contravariant object reads $\p^\mu = h^{\mu\nu} \p_\nu= (0,\p_x,\p_y,\p_z)$. We will be using these Galilean derivatives throughout the rest of this work, and they are not to be confused with usual derivatives used in section \ref{sec2} for example.

\subsection{Galilean Isometries}
As mentioned in the introduction, GCA is a Galilean or Non-Relativistic contraction of the relativistic conformal algebra. Equivalently to the intrinsic description in the last section, Galilean Conformal objects can be realised by taking the following contraction of coordinates on associated conformal theories  \cite{Bagchi:2009my}:
\bea{}\label{contraction} x_i \rightarrow \e x_i, t\rightarrow t, \e \rightarrow 0.
\eea
The above turns out to be equivalent to taking a $c\to \infty$ scaling.
Remember that in $4d$ the conformal algebra is a finite dimensional algebra, and hence to start with we only get the finite part of the GCA (fGCA) when the above mentioned contraction is acted upon. This finite algebra is generated by rotations ($J_{ij}$), spacetime translations ($H$ and $P_i$), boosts ($B_i$), scaling ($D$) and special conformal transformations ($K$ and $K_i$). The vector fields associated to these generators are given by:
\bea{}\label{gp}
&J_{ij} = - (x_i \p_j - x_j \p_i), \qquad
P_i = \p_i, \qquad H = -\p_t, \qquad
B_i = t\p_i\\
&D = - (t\p_t + x^i\p_i), \qquad
K = - (2tx^i\p_i + t^2\p_t), \qquad K_i = t^2\p_i.
\eea
The $i,j$ indices all correspond to purely spatial components in the above.
Consider an extension of the generators in an $n$ dependent form
\be{}\label{fs}
L^{(n)} = -(n+1)t^n x^i \p_i - t^{n+1}\p_t , \qquad M_i^{(n)} = t^{n+1} \p_i ,
\ee
where for $n=0,\pm 1$, the generators $L^{(n)},M^{(n)}_{i}$ denotes the set of finite GCA generators
\bea{}\label{f3}
L^{(-1,0,1)} = \{H, D, K\},~~ M_i^{(-1,0,1)} = \{P_i , B_i , K_i\} ,
\eea
but in principle any value of $n$ is admissible, hence giving rise to generators spanning an infinite dimensional vector space. The rotation generators could also be given an infinite dimensional lift as follows:
\be{}
J_{ij}^{(n)} = -t^n (x_i\p_j - x_j \p_i).
\ee
Armed with these new generators the full infinite dimensional extended GCA can  be written in the following form:
\bea{}\label{algebra}
&&\lb L^{(n)} , L^{(m)} \rb = (n-m)L^{(n+m)},~ \lb L^{(n)}, M_i^{(m)} \rb = (n-m)M_i^{(n+m)},\lb M_i^{(n)} , M_j^{(m)} \rb = 0,\nonumber\\
&& \lb L^{(n)}, J_{ij}^{(m)} \rb = -m J_{ij}^{(n+m)},~
\lb J_{ij}^{(n)}, M_r^{(m)} \rb = -(M_i^{(n+m)} \delta_{jr} - M_j^{(n+m)} \delta_{ir}), \nonumber\\&&
\lb J_{ij}^{(n)} , J_{rs}^{(m)} \rb = \delta_{is} J_{rj}^{(n+m)} + \delta_{jr} J_{si}^{(n+m)} + \delta_{ir} J_{js}^{(n+m)} + \delta_{js} J_{ir}^{(n+m)}.
\eea
In the rest of the paper, we will be talking about  theories which are manifestly invariant under transformations induced by this algebra.

\subsection{Covariant Lagrangian and transformation laws}
Conventionally, Galilean electrodynamics, and more specifically the conformal cousin of it, has been studied in the literature from an equation of motion point of view \cite{LBLL,Bagchi:2014ysa}. This hinges on the fact that there can be two distinct limits of relativistic Electrodynamics, known as the Electric and Magnetic ones, that may correspond to a theory of Galilean Electrodynamics. In the electric limit, the time-like components of the gauge field $A^\mu$ dominate (i.e. $ E_i \gg B_i$), while in the magnetic case the space-like components of the same dominate (i.e. $B_i \gg E_i$). For a theory with sources, one could take the same limits on the currents to write electric  and magnetic equations of motion. To this effect, in \cite{Mehra:2021sfx} a composite albeit covariant Lagrangian was  introduced, which consistently reproduces both electric and magnetic equations of motion.

The source free action for the Galilean Covariant Lagrangian can be written as:
\bea{4.2}\label{gali}
S(a_{\mu}, a^{\mu}, \p_{\mu}a_{\nu}, \p^{\mu}a^{\nu})= \int d^3x\,dt\,\Big[-\frac{1}{4}f_{\mu\nu}f^{\mu\nu}\Big],
\eea
where the contravariant and covariant field strengths are given in terms of Galilean gauge field $a^\mu$ and $a_\mu$. These fields here are distinct objects due to the structure of Galilean tensors, and the respective field strengths read:
\be{}
f^{\mu\nu} = (\p^\mu a^\nu - \p^\nu a^\mu),~~f_{\mu\nu} = (\p_\mu a_\nu - \p_\nu a_\mu).
\ee
So there are two distinct equations of motion, obtained by varying the above action w.r.t. one of the two kinds of gauge fields, which do not depend on each other. One can see while variation w.r.t. $a^\mu$ gives rise to,
\be{sds}
\p_{\nu} f^{\mu\nu} =0,~
\ee
which are the equations of motion in the Electric limit, on the other hand,
variation w.r.t $a_\mu$ leads one to 
\be{}
\p^{\nu} f_{\mu\nu} = 0,
\ee
generating the magnetic equations of motion \footnote{In component form, the Electric and Magnetic equations can be written as 
\bea{EMEQ}
&&
\p_i\p^i a^0 = 0,~
\p^j (\p_t a^0+\p_i a^i)=(\p_i\p^i) a^j ~~~\text{(Electric)}\\&& \nonumber
\p_t \p_i a_i=\p_i\p^i a_0 ,~
\p^j \p_i a_i=\p_i\p^i a_j ~~~\text{(Magnetic)}. 
\eea} (see \cite{Bagchi:2014ysa} for details). So evidently, the contravariant gauge fields are responsible for the electric limit of the theory, while covariant ones are responsible for the magnetic limit of the theory\footnote{Covariant gauge fields are defined by $a^\mu \t_\mu =0$ and contravariant ones are defined by $a_\mu h^{\mu\nu}= 0$, these belong to invariant vector spaces under the action of the Galilean group.}.

We will now write down the transformation of both kinds of the gauge fields under GCA\footnote{To get a better understanding of the representation theory, we urge the reader to look at \cite{Bagchi:2014ysa,Mehra:2021sfx}.}, which will be important in the two separate limits. The covariant formalism used here ensures that the theory is invariant under Galilean boosts, rotations and translations. For gauge fields in magnetic limit, the transformation laws for the covariant fields under rotations, Galilean boost, scale transformations and the special conformal transformations (SCT) take the following form:
\underline{Rotations}:
\bea{}&&
\delta_{J}a_0 =(x^{i}\p^{j}-x^{j}\p^{i})a_0,~~\delta_{J}a_k =(x^{i}\p^{j}-x^{j}\p^{i})a_k+(\delta^{i}_{k}a_j-\delta^{j}_{k}a_i).
\eea
\underline{Galilean boosts}:
\bea{}&&
\delta_{B_m}a_0 =-t\p_m a_0 - a_m,~~\delta_{B_m}a_i =-t\p_m a_i.
\eea
\underline{Scale transformations}:
\bea{}\delta_{D} (a_0,a_i )=(t\p_t+x^l\p_l+1) (a_0,a_i ),
\eea
\underline{SCT}:
\bes{}
\bea{}&&\delta_{K} a_0=(t^{2}\p_t+2t x^l\p_l+2t)a_0 +2 x^l a_l,~~\delta_{K} a_i=(t^{2}\p_t+2t x^l\p_l+2t)a_i,\\&&
\delta_{K_m}a_0 =-t^{2}\p_m a_0 -2t a_m,~~\delta_{K_m}a_i =-t^{2}\p_m a_i.
\eea\ees
Finally, we will write the variation of gauge fields in magnetic limit under infinite dimensional GCA generators $(L^{(n)},M^{(n)}_{m})$. They are given by
\bes{}
\bea{}&&\delta_{L^{(n)}} a_0=(t^{n+1}\p_t+(n+1)t^n x^l\p_l+(n+1)t^n)a_0 +n(n+1)t^{n-1}x^l a_l,\\&&
\delta_{L^{(n)}} a_i=(t^{n+1}\p_t+(n+1)t^n x^l\p_l+(n+1)t^n)a_i,\\&&
\delta_{M^{(n)}_m}a_0 =-t^{n+1}\p_m a_0 -(n+1)t^n a_m,\\&&
\delta_{M^{(n)}_m}a_i =-t^{n+1}\p_m a_i.
\eea\ees
Similarly, in the electric limit, we have transformation laws for contravariant gauge fields:\\
\underline{Rotations}:
\bea{}&&
\delta_{J}a^0 =(x^{i}\p^{j}-x^{j}\p^{i})a^0,~~\delta_{J}a^k =(x^{i}\p^{j}-x^{j}\p^{i})a^k+(\delta^{ki}a^{j}-\delta^{kj}a^i).
\eea
\underline{Galilean boosts}:
\bea{} \delta_{B_m}a^0 =-t\p_m a^0,~~ \delta_{B_m}a^i =-t\p_m a^i+ \delta^{i}_{m}a^0.\eea
\underline{Scale transformations}:
\bea{}\delta_{D} (a^0,a^i )=(t\p_t+x^l\p_l+1) (a^0,a^i ).
\eea
\underline{SCT}:
\bes{}
\bea{}&&\delta_{K} a^0=(t^{2}\p_t+2t x^l\p_l+2t)a^0,~~\delta_{K} a^i=(t^{2}\p_t+2t x^l\p_l+2t)a^i-2x^i a^0,\\&&
\delta_{K_m}a^0 =-t^{2}\p_m a^0,~~\delta_{K_m}a^i =-t^{2}\p_m a^i+2t \delta^{i}_{m}a^0.
\eea\ees
Under $(L^{(n)},M^{(n)}_{m})$, the transformation laws are given by
\bes{}
\bea{}&&\delta_{L^{(n)}} a^0=(t^{n+1}\p_t+(n+1)t^n x^l\p_l+(n+1)t^n)a^0,\\&&
\delta_{L^{(n)}} a^i=(t^{n+1}\p_t+(n+1)t^n x^l\p_l+(n+1)t^n)a^i-n(n+1)t^{n-1}x^i a^0,\\&&
\delta_{M^{(n)}_m}a^0 =-t^{n+1}\p_m a^0,\\&&
\delta_{M^{(n)}_m}a^i =-t^{n+1}\p_m a^i+(n+1)t^n \delta^{i}_{m}a^0.
\eea\ees

One can easily see the sheer asymmetry between the transformations of covariant and contravariant objects in this case, and of course the same shows up between temporal and spatial components.
Using these above transformations, one could deduce the relevant transformation laws for the electric and magnetic field strengths as well, and explicitly check the invariance of \eqref{gali} under the same.

\subsection{Electric and Magnetic invariants}
We have seen earlier that a relativistic ModMax Lagrangian depends on both Lorentz invariants in Electrodynamics.
Hence for the purpose of this work, defining Electromagnetic invariants under Galilean transformations are very important.
Now for example, $f^{\mu\nu}$ is clearly an Electric object, since the gauge fields are contravariant here, similarly $f_{\mu\nu} $ is a magnetic object for similar reasons.  To mark their properties, we call them  $f^{\mu\nu}_{(E)}$ and $f_{\mu\nu}^{(M)} $  from now and hereon. 

The obvious invariant quantity is the covariant Lagrangian for Galilean Maxwell theory, which is a composite of Electric and Magnetic objects,
\be{trL}
\L  = -\frac{1}{4}\fE\fM,
\ee
i.e. a ``true'' Lagrangian is one composed of both Electric and Magnetic tensors with contracted indices, and gives the right Electric or Magnetic EOM when varied w.r.t one or the other gauge fields. But this is not the Lagrangian when one takes the relativistic Lagrangian and performs an Electric or a Magnetic limit. In that case, both field strength components of the Lagrangian change, i.e. we get either of
\be{pck}
\L^{(E)} = -\frac{1}{4}\fE\fEE, ~~\L^{(M)} = -\frac{1}{4}\fMM\fM,
\ee
which are only useful in one or the other limits.
Here, the inverse field strengths $\fEE$ and $\fMM$ are not electric or magnetic objects respectively, not at least by contra/co-variance of the associated gauge field. But they are \textit{electric inverse of the electric field strength} and \textit{magnetic inverse of the magnetic field strength}, in the same vein as defining the projective inverses for our Galilean tensors. To connect to the notation of \cite{Mehra:2021sfx}, these are actually defined by the tilde conjugation, which acts via a Galilean contraction of gauge fields. For example, in the Electric case,
\bea{}
\fEE=\tilde{f}_{\mu\nu} =- f^{\a\b} T_{\mu\a\b\nu} = (\p_\mu \tilde{a}_\nu - \p_\nu \tilde{a}_\mu),~
\tilde{a}_\mu = a^\nu \t_{\mu\nu},
\eea
Where the tensor $T$ is defined as a combination of $\t$ and $h$ to achieve this,
\bea{T}
T_{\a\b\mu\nu} := 4 \t_{[\a} h_{\b] [\mu} \t_{\nu]} = (\t_\a h_{\b\mu} \t_\nu - \t_\b h_{\a\mu} \t_\nu - \t_\a h_{\b\nu} \t_\mu + \t_\b h_{\a\nu} \t_\mu).
\eea
It is easy to see that $T_{\a\b\mu\nu}$ is symmetric under exchange of $\a$ and $\nu$ and of $\b$ and $\mu$ i.e.
$T_{\a\b\mu\nu} = T_{\nu\mu\b\a}$.
Notice that $T_{\a\b\mu\nu}$ is antisymmetric if we exchange $\a$ with $\b$ or $\mu$ with $\nu$:
\be{}
T_{\a\b\mu\nu} = - T_{\b\a\mu\nu} = - T_{\a\b\nu\mu} = T_{\b\a\nu\mu}.
\ee
From these it also follows that $T_{\a\b\mu\nu} - T_{\a\mu\b\nu} = T_{\a\nu\mu\b}$.
Similarly the magnetic inverse is given by contraction with only $h$'s,
\be{}
\fMM=\tilde{f}^{\mu\nu} = h^{\a\mu}f_{\a\b}h^{\b\nu} =  (\p^\mu \tilde{a}^\nu - \p^\nu \tilde{a}^\mu), ~~\tilde{a}^\mu = h^{\mu\nu}a_\nu.
\ee
Note, one of these field strengths are dualised by $\t_{\mu\nu}$ and the other by $h^{\mu\nu}$, thereby giving them the notion of an Electric (temporal) or a Magnetic (spatial) contraction. Hence the Lagrangians \eqref{pck} $\L^{(E/M)}$ are not ``true'' Lagrangians, but limits of relativistic Lagrangians in the respective regimes where only notion of Electric terms or Magnetic terms survive.  By definition,  $\L^{(E/M)}$ both are Galilean invariants as one can explicitly show, and we will go ahead to define an additional GCA invariant other than that of $\mathcal{L}$ in \eqref{trL}:
\be{inv}
\mathcal{M}=\frac{1}{\sqrt{2}}\sqrt{\mathcal{L}^{(E)~2}+\mathcal{L}^{(M)~2}}.
\ee
This particular quantity will be crucial in our later discussions. 

Although we have defined a notion of ``dual'' field strengths for Galilean theories using the tilde conjugation, a real Hodge dual in this case is ill-defined as the associated metric degenerates. Evidently, the notion of EM duality is lost as the two regimes aren't simply connected in a Galilean theory. To the best of our knowledge, the notion of Hodge duals on a Newton-Cartan manifold is not discussed in the literature as well. However, we can always go ahead and define the Hodge-dual-like field strength tensor for the Galilean case in accordance with its relativistic counterpart: $\star f_{\mu\nu} =\frac{1}{2}\e_{\mu\nu\rho\sigma}f^{\rho\sigma}$, 
assuming the definition of Levi-Civita will not change under NR limits. In this sense it relates true electric and magnetic representations on either sides of the equality
\be{}
\star f_{\mu\nu}^{(M)} = \frac{1}{2}\e_{\mu\nu\rho\sigma}f^{\rho\sigma}_{(E)}~~\text{and}~~\star f^{\mu\nu}_{(E)}=\frac{1}{2}\e^{\mu\nu\rho\sigma}f_{\rho\sigma}^{(M)}.
\ee
As we discussed before, a true contracted Galilean object will be a combination of purely Electric and Magnetic tensors. And based on the definitions we provided earlier, we can show that
\be{}
\L^{*(E)}  = -\frac{1}{4}\fE\star f_{\mu\nu}^{(E)}~~\text{or}~~\L^{*(M)}  = -\frac{1}{4}\fM\star f^{\mu\nu}_{(M)},
\ee
are both invariant under the GCA transformations as well. 
Here we have gone further to define the \textit{Electric and Magnetic inverses of the Hodge dual tensor}:
\be{}
\star f_{\mu\nu}^{(E)} =  \frac{1}{2}\e_{\mu\nu\rho\sigma}f^{\rho\sigma}_{(M)} =  \frac{1}{2}\e_{\mu\nu\rho\sigma}\tilde{f}^{\rho\sigma},~~~
\star f^{\mu\nu}_{(M)} =   \frac{1}{2}\e^{\mu\nu\rho\sigma}f_{\rho\sigma}^{(E)} = \frac{1}{2}\e_{\mu\nu\rho\sigma}\tilde{f}_{\rho\sigma} .
\ee
We should again remind the reader, the objects defined under this star conjugation are not real Galilean objects, and are only defined in an ad-hoc basis. We will comment on these invariants later in the paper. 

\section{Galilean Conformal ModMax-like Lagrangian}\label{sec4}
\subsection{Symmetries of the Lagrangian}
Let us now come to the crux of this paper, that is to construct a non-linear ModMax-like, albeit GCA invariant, Galilean electrodynamics theory. As in the case of the Relativistic ModMax theory, we can define a Lagrangian in terms of the two GCA invariants $\mathcal{L}$ and $\mathcal{M}$ we described in the last section (see \eqref{trL} and \eqref{inv}), with the same general distinctive structure:
\bea{}\label{Gmm6}
\mathcal{L}_{GMM}&=&  -\frac{\cosh\gamma}{4}~\mathcal{L} + \frac{\sinh\gamma}{4} ~\sqrt{\mathcal{L}^2+\mathcal{M}^2}\\ \nonumber
&=& -\frac{\text{cosh}\gamma}{4}\big[f^{\mu\nu}f_{\mu\nu}\big]+\frac{\text{sinh}\gamma}{4}\sqrt{(f^{\mu\nu}f_{\mu\nu})^2+\frac{1}{2}(\tilde{f}^{\mu\nu}f_{\mu\nu})^2+\frac{1}{2}(\tilde{f}_{\mu\nu}f^{\mu\nu})^2}.
\eea
This simply becomes \eqref{gali} when we choose $\gamma=0$.
In component form, the Lagrangian can be written down as
\bea{}\label{GMM5}\mathcal{L}_{GMM}=-\frac{\text{cosh}\gamma}{4}E+\frac{\text{sinh}\gamma}{4}\sqrt{C},\eea
where written in component form, the quantities read: $C=[E^2+\frac{1}{2}(\tilde{f}^{ij}f_{ij})^2+\frac{1}{2}(2\tilde{f}_{i0}(\p^{i} a^{0}))^2]$ and $E=[2f_{i0}(\p^{i} a^{0})+f^{ij}f_{ij}]$.
Let us now move on to the invariance of this Lagrangian \eqref{GMM5} under GCA. The Lagrangian is trivially invariant under translations and rotations. We will only show the invariance under the boost, scale transformations and SCT. The change of the Lagrangian under action of the Boost generator is given by
\bea{}&&\hspace{-1cm}\delta_{B}\mathcal{L}_{GMM}=\frac{\text{cosh}\gamma}{4}\p_m[t(2f_{i0}(\p^{i} a^{0})+f^{ij}f_{ij})]-\frac{\text{sinh}\gamma}{4}\frac{1}{2\sqrt{C}}\p_{m}(t C)=\p_m(-t\mathcal{L}_{GMM}).
\eea
We will next check the variation under scale transformation. It is given by
\bea{}&&\delta_D\mathcal{L}_{GMM}=\frac{\text{cosh}\gamma}{4}[(t\p_t+x^m\p_m+4)E]+\frac{\text{sinh}\gamma}{4}[(t\p_t+x^m\p_m+4)\sqrt{C}]\nonumber\\&&\hspace{1.7cm}=(t\p_t+x^m\p_m+4)\mathcal{L}_{GMM}=\p_t(t\mathcal{L}_{GMM})+\p_m(x^m\mathcal{L}_{GMM}).\eea
Finally looking at the change of \eqref{GMM5} under Special Conformal Transformations $K$, we get
\bea{}&&\delta_K\mathcal{L}_{GMM}=(t^2\p_t+2tx^m\p_m+8t)\Big[-\frac{\text{cosh}\gamma}{4}E+\frac{\text{sinh}\gamma}{4}\sqrt{C}\Big]\nonumber\\&&\hspace{1.7cm}=\p_t(t^2\mathcal{L}_{GMM})+\p_m(2tx^m\mathcal{L}_{GMM}).\eea
We have looked at the invariance of \eqref{GMM5} under finite part of GCA. Next step will be to move on to the infinite extension of GCA. Under the generators $M^{(n)}_{m}$, we get the variation:
\bea{}&&\hspace{-1cm}\delta_{M}\mathcal{L}_{GMM}=\frac{\text{cosh}\gamma}{4}\p_m(t^{n+1}E)-\frac{\text{sinh}\gamma}{4}(t^{n+1}\p_{m} \sqrt{C})=\p_m(-t^{n+1}\mathcal{L}_{GMM}).
\eea
Similarly, under $L^{(n)}$, we have,
\bea{}&&\delta_L\mathcal{L}_{GMM}=(t^{n+1}\p_t+(n+1)t^n x^m\p_m+4(n+1)t^n)\Big[-\frac{\text{cosh}\gamma}{4}E+\frac{\text{sinh}\gamma}{4}\sqrt{C}\Big]\nonumber\\&&\hspace{1.7cm}=\p_t(t^{n+1}\mathcal{L}_{GMM})+(n+1)\p_m(t^n x^m\mathcal{L}_{GMM}).\eea
In all of these cases, the transformations change the Lagrangian via a total derivative term, and hence we see that the theory is fully invariant under extended part of GCA.

Now few comments are in order at this point. Clearly the structure of \eqref{Gmm6}, like the relativistic counterpart, depends on the use of electromagnetic invariants, which occur directly in the Lagrangian. In the relativistic case, choice of these invariants are straightforward, however it is evidently not so simple in the Galilean counterpart, as we have discussed before. It actually turns out that instead of $\mathcal{M}$ one could choose some other Galilean invariant as mentioned in \ref{}. Consequently one may go ahead and write down a test Lagrangian of the form:
\be{}
\mathcal{\bar{L}}_{GMM}=-\frac{\text{cosh}\gamma}{4}[f^{\mu\nu}_{(E)}f_{\mu\nu}^{(M)}]+\frac{\text{sinh}\gamma}{4}\sqrt{(f^{\mu\nu}_{(E)}f_{\mu\nu}^{(M)})^2+(\fE\star f_{\mu\nu}^{(E)})^2}
\ee
or an equivalent one with the $f\star f$ term replaced by the magnetic version. It can be explicitly shown using methods we discussed in this section that the above Lagrangian is invariant under GCA symmetries as well. However as we mentioned earlier, these $f\star f$ terms in the Galilean case are defined in an ad-hoc way, hence we do not delve into detail discussions on them. For us, \eqref{Gmm6} will be the master Lagrangian to follow through. 
\subsection{Equations of motion and Gauge invariance}

The equations of motion from \eqref{Gmm6} comes out to be twofold, as in the case of its Maxwellian cousin. Varying the action w.r.t. the contravariant fields $a^\mu$, we get the Electric-like equation of motion:
\be{}
\p_{\mu}\Bigg[(\text{cosh}\gamma)f^{\mu\nu}-\text{sinh}\gamma\Bigg(\frac{(f^{\a\b}f_{\a\b})f^{\mu\nu}+(\tilde{f}^{\a\b}f_{\a\b})\tilde{f}^{\mu\nu}}{\sqrt{(f^{\a\b}f_{\a\b})^2+\frac{1}{2}(\tilde{f}^{\a\b}f_{\a\b})^2+\frac{1}{2}(\tilde{f}_{\a\b}f^{\a\b})^2}}\Bigg)\Bigg]=0,
\ee
while varying w.r.t. covariant gauge fields $a_\mu$ gives us the Magnetic-like equation of motion
\be{}
\p^{\mu}\Bigg[(\text{cosh}\gamma)f_{\mu\nu}-\text{sinh}\gamma\Bigg(\frac{(f^{\a\b}f_{\a\b})f_{\mu\nu}+(\tilde{f}_{\a\b}f^{\a\b})\tilde{f}_{\mu\nu}}{\sqrt{(f^{\a\b}f_{\a\b})^2+\frac{1}{2}(\tilde{f}^{\a\b}f_{\a\b})^2+\frac{1}{2}(\tilde{f}_{\a\b}f^{\a\b})^2}}\Bigg)\Bigg]=0.
\ee
Observe that both equations reduce down to the Galilean Maxwell Electric and Magnetic equations of motion when we put in $\gamma=0$. Also these are in the same footing as the equations of motion from Relativistic version of the ModMax theory.  One can look at these equations and note that the electric and the magnetic sectors are interchanged via the exchange of field tensors $f^{\mu\nu}  \leftrightarrow f_{\mu\nu}$ (which is equivalent to doing $E_i \to B_i$ and $B_i \to - E_i$ in the relativistic case). One can think of this as a Galilean remnant of the original electromagnetic duality. 

We will now look at the gauge transformations (GT) for this theory. The gauge transformations for $a^{\mu}$ and $a_{\mu}$, that keep the structure of Lagrangian unchanged, are given by
\bes{}\label{gi}
\bea{} &&a^{\mu}\rightarrow a^{\mu}+\p^{\mu}\Lambda_1 \implies f^{\mu\nu} \rightarrow f^{\mu\nu}\\&&
a_{\mu}\rightarrow a_{\mu}+\p_{\mu}\Lambda_2 \implies f_{\mu\nu} \rightarrow f_{\mu\nu}
\eea\ees
whereas $(\Lambda_{1,2})$ are two different gauge potentials corresponding to symmetries in either limit. \footnote{One would use an analogue of the Lorenz gauge in either limit i.e. $\p_\mu a^\mu =0$ or $\p^\mu a_\mu =0$, and this would imply the gauge potentials satisfy a Laplace equation $\nabla^2 \Lambda_{1,2}=0$.}
 Similarly, the transformation of the conjugate field strenghts $\tilde{f}^{\mu\nu}$ and $\tilde{f}_{\mu\nu}$ under \eqref{gi} are given by
\bea{}\label{git}\delta_{GT}\tilde{f}^{\a\b}=h^{\mu\a}(\delta_{GT}f_{\mu\nu})h^{\nu\b}=0,~~\delta_{GT}\tilde{f}_{\a\b}=-(\delta_{GT}f^{\mu\nu})T_{\a\mu\nu\b}=0.\eea
We this conclude that the Lagrangian and equations of motion are invariant under gauge transformations \eqref{gi}.

\subsection{Energy-momentum tensors}
Let us now write down the energy-momentum (EM) tensors in the Electric and Magnetic limit of Galilean ModMax theory \eqref{Gmm6}. We will use the Noether charge methodology followed in \cite{Mehra:2021sfx} to deduce purely electric or magnetic stress tensors in either of those limits. In the Electric limit, it is given by
\be{EEM}
{T_E}^\mu_{\;\;\nu} = \Big( f^{\mu\a}\tilde{f}_{\a\nu} + \frac{1}{4} \delta^\mu_\nu f^{\a\b}\tilde{f}_{\a\b}\Big)\Big( \cosh\gamma-\sinh\gamma \frac{\L^{(E)}}{\sqrt{\mathcal{L}^2+\mathcal{M}^2}}\Big),
\ee
whereas in the Magnetic limit, the stress tensor reads
\be{MEE}
{T_M}^\mu_{\;\;\nu} = \Big( \tilde{f}^{\mu\a}f_{\a\nu} + \frac{1}{4}\delta^\mu_\nu \tilde{f}^{\a\b}f_{\a\b}\Big)\Big( \cosh\gamma-\sinh\gamma \frac{\L^{(M)}}{\sqrt{\mathcal{L}^2+\mathcal{M}^2}}\Big).
\ee
This is again reminiscent of the relativistic ModMax case, as our stress tensors in either limits are explicitly proportional to the Galilean Maxwell ones of \cite{Mehra:2021sfx}, with a multiplied contracted term. 
As we know for generic Galilean Conformal theories, the stress tensor needs to be traceless i.e. ${T}^\mu_{\;\;\mu} =0$ and the condition on the component $T^{0}_{~i}=0$ has to be satisfied, since there is no momentum flux in non-relativistic theories \cite{deBoer:2017ing}. In case of the Electric limit, it is easy to check these conditions:
  \bea{EEM1}&&
{T_E}^\mu_{\;\;\mu} = \Big( -f^{\a\mu}\tilde{f}_{\a\mu} +  f^{\a\b}\tilde{f}_{\a\b}\Big)\Big( \cosh\gamma-\sinh\gamma \frac{\L^{(E)}}{\sqrt{\mathcal{L}^2+\mathcal{M}^2}}\Big)=0,\\&&
{T_E}^0_{~i} = \Big( f^{0 \a}\tilde{f}_{\a i} \Big)\Big( \cosh\gamma-\sinh\gamma \frac{\L^{(E)}}{\sqrt{\mathcal{L}^2+\mathcal{M}^2}}\Big)=0.
\eea
Similarly, in the Magnetic limit, the stress tensor satisfies 
\bea{MEE}&&
{T_M}^\mu_{\;\;\mu} = 0,~{T_M}^0_{~i} = 0.
\eea
This shows both electric and magnetic sectors of our non-linear theory is explicitly Galilean invariant as well.

\section{Discussions and conclusions}\label{sec5}
In this short paper, we described a nonlinear Galilean Covariant Lagrangian that is invariant under Galilean Conformal symmetries by construction. Interestingly, the Lagrangian was written in the same vein of the ModMax Lagrangian and hence reaffirms the conformal nature of the ModMax construction beyond the relativistic case.  We focussed on the invariants of the Galilean Maxwell theory and used them as building blocks to build our Lagrangian, with an explicit proof of invariance under GCA transformations. We also discussed the Galilean equations of motion and stress tensors, both in the Electric and Magnetic limits of the theory. The nonlinear equations in all of the cases reduce to the known Galilean equations in the $\gamma=0$ limit. 

As we mentioned earlier, our calculation here introduces the first example of a nonlinear Galilean covariant electrodynamics theory. It remains to explore whether the usual NLED physics, like classical solutions, carry forward to these sort of Galilean theories as well. As we mentioned earlier, it is useful to go into the Hamiltonian formalism for the relativistic ModMax theory in order to make sense of null solutions. It would be interesting to explore the canonical structure of the Galilean theory as well along this route, starting from \eqref{gali} and proceeding to find the same for the total Galilean ModMax theory.  Another very straightforward extension would be to discuss Galilean superconformal extension of this theory in the vein of \cite{Bandos:2021rqy}. Super-GCA algebras have already been described well in the literature \cite{Sakaguchi:2009de, Bagchi:2009ke, deAzcarraga:2009ch} and one would hope to find a nonlinear realisation of that as well in the super-ModMax-like theory. We hope to come back to these issues in future work.

\section*{Acknowledgements}
The authors would like to thank Arjun Bagchi and Rudranil Basu for useful discussions and comments on the manuscript.
The work of ArB is supported by the Quantum Gravity Unit of the Okinawa Institute of Science and Technology Graduate University (OIST). AM acknowledges support of the Department of Physics at BITS Goa through the grant SERB CRG/2020/002035 and School of Mathematics at University of Edinburgh where part of this work was conducted. ArB acknowledges kind hospitality from IIT Kanpur and Kyoto University during the course of this work. 
\bibliographystyle{utphysmodb}
\bibliography{ref.bib}

\end{document}